\documentclass{ws-p8-50x6-00}
\newcommand{\eqref}[1]{(\ref{#1})}

\begin{document}

\def\deg{\hbox{$^\circ$}}
\def\sun{\hbox{$\odot$}}
\def\earth{\hbox{$\oplus$}}
\def\la{\mathrel{\hbox{\rlap{\hbox{\lower4pt\hbox{$\sim$}}}\hbox{$<$}}}}
\def\ga{\mathrel{\hbox{\rlap{\hbox{\lower4pt\hbox{$\sim$}}}\hbox{$>$}}}}

\title{The magnetic stress tensor\\ in magnetized matter}

\author{Olivier Espinosa}

\address{Departamento de F\'{\i}sica,
Universidad T\'{e}cnica Federico Santa Mar\'{\i}a,\\
Valpara\'{\i}so, Chile
\\ olivier.espinosa@fis.utfsm.cl}

\author{Andreas Reisenegger}

\address{Departamento de Astronom\'{\i}a y Astrof\'{\i}sica, Facultad de
F\'{\i}sica,\\
Pontificia Universidad Cat\'{o}lica de Chile, Santiago, Chile
\\ areisene@astro.puc.cl}


\maketitle

\centerline{To ${B^2\over 8\pi}$ or not to ${B^2\over 8\pi}$, {\it
that} is the question!}

\centerline{\small\it(adapted from J. Horvath's Concluding
Remarks, with apologies to W. Shakespeare)}

\bigskip

\abstracts{We derive the form of the magnetic stress tensor in a
completely general, stationary magnetic medium, with an arbitrary
magnetization field $\vec M(\vec r)$ and free current density
$\vec j(\vec r)$. We start with the magnetic force density $\vec
f$ acting on a matter element, modelled as a collection of
microscopic magnetic dipoles in addition to the free currents. We
show that there is a unique tensor ${\bf T}$ quadratic in the
magnetic flux density $\vec B(\vec r)$ and the magnetic field
$\vec H(\vec r)=\vec B-4\pi\vec M$ whose divergence is
$\nabla\cdot{\bf T}=\vec f$. In the limit $\vec M=0$, the
well-known vacuum magnetic stress tensor is recovered. However,
the general form of the tensor is asymmetric, leading to a
divergent angular acceleration for matter elements of vanishing
size. We argue that this is not inconsistent, because it occurs
only if $\vec M$ and $\vec B$ are not parallel, in which case the
macroscopic field does indeed exert a torque on each of the
microscopic dipoles, so this state is only possible if there are
material stresses which keep the dipoles aligned with each other
and misaligned with the macroscopic field. We briefly discuss the
consequences for the stability of strongly magnetized stars.}

\section{Introduction}

The equilibrium thermodynamics of magnetized matter is important
to understand the structure and evolution of compact astrophysical
objects such as white dwarfs and neutron stars. Of particular
relevance is the pressure, since it enters in an important way in
the equations that determine the structure of the star (Newtonian
hydrostatic equilibrium equation or its relativistic
Oppenheimer-Volkov generalization). If the gas that conforms the
star is made of electromagnetically active matter (i.e., ionized
matter or magnetic dipoles) in the presence of macroscopic
electromagnetic fields, the total pressure will have both a
kinetic and an electromagnetic component.

There seems to be some confusion and disagreement about the nature
of the latter (and also about the isotropy of the total pressure).
For instance, recent
papers\cite{Chaichian,Herman+Hugo+Aurora+otro,Herman+Hugo+Aurora}
claim that, for a magnetization parallel to the magnetic field,
the electromagnetic contribution to the pressure takes the form
\begin{equation}\label{electromagnetic pressure-Herman}
P_\parallel = 0,\qquad P_\perp = -MB,
\end{equation}
\noindent
where $\parallel$ and $\perp$ refer to the direction of the
magnetic field.
The authors of reference\cite{Herman+Hugo+Aurora} claim that, with
an electromagnetic pressure of the this form
and some assumptions about the growth of the
magnetic field from the surface of the star towards its center,
magnetars with surface fields of the order of $10^{15}$ Gauss
would be unstable to collapse.

However, the correct electromagnetic pressure in
the case of no magnetization, as derived from the Maxwell stress
tensor for fields in vacuum is
\begin{equation}\label{electromagnetic pressure-Maxwell}
P_\parallel^{\rm vac} = -\frac{B^2}{8\pi},
\qquad P_\perp^{\rm vac} = \frac{B^2}{8\pi},
\end{equation}
\noindent
which clearly does not correspond to the zero magnetization limit
of expressions (\ref{electromagnetic pressure-Herman}).


In this work we address this problem by studying the
electromagnetic force on a piece of magnetized matter from a
purely classical point of view and constructing thereby the
corresponding magnetic stress tensor for the system. Our results
imply that the electromagnetic contribution to the pressure,
defined as the negative of the diagonal components of the stress
tensor, in the case of parallel magnetization and magnetic field
is given by
\begin{equation}\label{electromagnetic pressure-ours}
P_\parallel = -\frac{B^2}{8\pi},\qquad P_\perp = \frac{B^2}{8\pi}-MB,
\end{equation}
which has the correct vacuum limit \eqref{electromagnetic
pressure-Maxwell} and, naturally, disagrees with \eqref{electromagnetic
pressure-Herman}.

Other works where the problem of the electromagnetic pressure has
been touched upon are\cite{Canuto,Easson,Blandford}.

\indent

\section{Magnetic force density on matter}

Macroscopic matter interacts with a magnetic field through its
electrical currents (due to motion of free charges) and
macroscopic magnetization (due to the alignment of microscopic
magnetic dipoles, usually associated with quantized spins).

The force density (force per unit volume) on the currents is
well-known to be $\vec f^{\rm curr}=\vec j\times\vec B/c$, where
$\vec j(\vec r)$ is the electric current density, and $\vec B(\vec
r)$ is the magnetic flux density.

The force on the dipoles can be found by considering the force on
a single dipole of moment $\vec m$, whose Cartesian components are
$F_i=m_j\nabla_iB_j$. Adding the dipole moments in a small volume,
we obtain the net force density, $f_i^{mag}=M_j\nabla_iB_j$, so
the total force density can be written as\cite{Pethick}
\begin{equation}\label{force density}
f_i={1\over c}\epsilon_{ijk}j_jB_k+M_j\nabla_iB_j,
\end{equation}
\noindent where $\epsilon_{ijk}$ are the components of the totally
antisymmetric (Levi-Civita) tensor, and the Einstein convention of
summation over repeated indices is understood here and in what
follows.

We emphasize that this force density is {\it not} the same as
obtained from defining a ``magnetization current'' $\vec
j^{mag}\equiv\nabla\times\vec M$ and writing $\vec f^{mag'}=\vec
j^{mag}\times\vec B/c$, although they agree when integrated over a
bounded material body. This is most clearly seen in the simple
case of a uniformly magnetized medium in a uniform magnetic field.
Our expression for $f_i^{mag}$ gives a uniform force per unit
volume, as physically expected from the uniform distribution of
dipoles acted on by the magnetic field. The alternative
expression, for $\vec f^{mag'}$, gives a vanishing force in the
interior of the body (zero ``magnetization current'' in the
interior), but a compensating force on the surface.

\section{The magnetic stress tensor}

\subsection{Derivation}

The component $T_{ij}$ of the stress tensor is generally defined
as minus the flux in the $j-$direction of the $i-$component of the
linear momentum. The force density $\vec f$ on a matter element is
the time derivative of its momentum density, i.e., minus the
divergence of the momentum flux, thus
\begin{equation}\label{force as divergence}
\vec f=\nabla\cdot{\bf T},\qquad{\rm or}\qquad f_i=\nabla_jT_{ij}.
\end{equation}
Using Maxwell's equations
\begin{equation}\label{maxwell equations}
\nabla\cdot\vec B=0\qquad{\rm and}\qquad \nabla\times\vec
H={4\pi\over c}\vec j,
\end{equation}
\noindent together with the definition $\vec H=\vec B-4\pi\vec M$,
eq. \eqref{force density} can be manipulated to write $\vec f$ as
the divergence of a possible stress tensor, with components
\begin{eqnarray}\label{our stress tensor}
T_{ij}&=&\frac{1}{4\pi}\left[H_iB_j-\left(\vec H\cdot\vec B-
\frac{1}{2}B^2\right)\delta_{ij}\right]\\
&=&\frac{B_iB_j}{4\pi}-\frac{B^2}{8\pi}\delta_{ij}-M_iB_j+\vec
M\cdot\vec B\delta_{ij}.
\end{eqnarray}
\subsection{Uniqueness?}

Of course, the form just obtained is not unique, since any tensor
with vanishing divergence can be added to ${\bf T}$ without
changing eq. \eqref{force as divergence}. Thus, we search for a
divergence-free tensor ${\bf\tilde T}$, quadratic in the magnetic
fields ($\vec B$ and $\vec H$),
whose general form is
\begin{equation}
\tilde
T_{ij}=\alpha_{ijkl}H_kH_l+\beta_{ijkl}H_kB_l+\gamma_{ijkl}B_kB_l,
\end{equation}
\noindent where $\alpha_{ijkl}$, $\beta_{ijkl}$, and
$\gamma_{ijkl}$ are 
components of constant tensors.
These tensors must be rotationally invariant (in order not to
introduce preferred directions other than those defined by the
magnetic fields), which constrains them to be linear combinations
of $\delta_{ij}\delta_{kl}$, $\delta_{ik}\delta_{jl}$, and
$\delta_{il}\delta_{jk}$.
Thus, we write
\begin{equation}
\tilde T_{ij}=(\alpha\vec H^2+\beta\vec H\cdot\vec B+\gamma\vec
B^2)\delta_{ij}+\alpha'H_iH_j+\beta'H_iB_j+\beta''B_iH_j+\gamma'B_iB_j,
\end{equation}
\noindent where $\alpha$, $\beta$, $\gamma$, $\alpha'$, $\beta'$,
$\beta''$, and $\gamma'$ are constant coefficients to be adjusted
so as to make the tensor divergence-free
for any choice of the vector fields $\vec H(\vec r)$ and $\vec
B(\vec r)$, with the only restriction that $\nabla\cdot\vec B=0$,
i.e., $\nabla_jB_j=0$.

In particular, we may first choose $\vec
B=0$ 
and $\vec H=x_a\hat e_b$ (i.e., $H_k=x_a\delta_{kb}$), where $a$
and $b$ are any {\it given} two different Cartesian indices, which
implies $\nabla_j\tilde T_{ij}=2\alpha x_a\delta_{ia}=0$ (no
summation over $a$), so $\alpha=0$. Changing $\vec H$ to be any
other vector field with $\nabla\cdot\vec H=0$ but $(\vec
H\cdot\nabla)\vec H\neq 0$ gives $\alpha'=0$. Exchanging the roles
of $\vec H$ and $\vec B$ in the arguments just given, we obtain
$\gamma=\gamma'=0$. Finally, putting $\vec H=x_a\hat e_b$ and
$\vec B=x_c\hat e_d$ with different combinations of values for
$a$, $b$, $c$, $d$ (respecting the constraint that $c\neq d$ to
satisfy $\nabla\cdot\vec B=0$) implies $\beta=\beta'=\beta''$.

Thus, 
among tensors quadratic in $\vec H$ and $\vec B$,
the one we originally obtained is unique in giving the correct
magnetic force density.

\subsection{Vacuum limit}

In the absence of magnetization,
the stress tensor given by eq. \eqref{our stress tensor} takes the
form
\begin{equation}
T_{ij}={B_iB_j\over 4\pi}-{B^2\over 8\pi}\delta_{ij},
\end{equation}
\noindent which is the correct, symmetric,
well-known\cite{Jackson} form in vacuum.

\subsection{Asymmetry and magnetic torque}

For arbitrary, non-parallel $\vec H$ and $\vec B$, the tensor is
manifestly non-symmetric, $T_{ij}\neq T_{ji}$. It is usually
argued\cite{Schutz} that the stress tensor must be symmetric, on
the ground that a non-symmetric stress tensor would produce a
torque on a small matter element which decreases less quickly with
decreasing volume than the moment of inertia of the element, thus
producing a divergent angular acceleration. We note that this
argument applies to the {\it total} stress tensor, which could be
composed of (in principle non-symmetric) pieces of different
physical origin.

In the present case, the asymmetry is present only if $\vec H$ and
$\vec B$ or, equivalently, $\vec M$ and $\vec B$ are not
collinear. This means that the microscopic dipoles are not aligned
with the macroscopic field, which therefore indeed causes a torque
on each dipole, which add up to a torque per unit volume
$\vec\tau=\vec M\times\vec B$. As a consistency check, we
calculate a component of the torque density in terms of the stress
tensor (as in Ref. \cite{Schutz}, but note the opposite sign
convention for the stress tensor),
$\tau_i=-\epsilon_{ijk}T_{jk}=-\epsilon_{ijk}H_jB_k/(4\pi)$, thus
$\vec\tau=-\vec H\times\vec B/(4\pi)=\vec M\times\vec B$,
completing the consistency check.

Of course, if there is no counter-acting torque (for example due
to microscopic interactions among neighboring dipoles), the
dipoles will orient themselves along $\vec B$, in which case the
stress tensor becomes symmetric and the torque disappears. If
there are microscopic interactions keeping the dipoles locally
aligned and at fixed positions with respect to each other, then a
macroscopic matter element may act as a rigid body with a finite
angular acceleration, that can not be subdivided into
infinitesimal pieces with a divergent angular acceleration,
because their magnetic torques are cancelled by the local
microscopic interactions.

\subsection{Magnetized fluid and anisotropic pressure}

In a stationary, magnetized fluid, there are no microscopic forces
keeping the the magnetization misaligned with the magnetic field,
therefore $\vec M(\vec r)$, $\vec B(\vec r)$, and $\vec H(\vec r)$
are collinear at every point $\vec r$.
Choosing a local Cartesian basis with $z-$axis aligned in the same
direction,
the stress tensor becomes diagonal, with components
$T_{xx}=T_{yy}=-P_\perp=MB-B^2/(8\pi)$ and
$T_{zz}=-P_\parallel=B^2/(8\pi)$. Thus, the parallel pressure
$P_\parallel<0$ always, corresponding to the usual tension along
field lines, while $P_\perp>0$ as usual, unless $M\geq B/(8\pi)$.

\section{Conclusions}

On the basis of physical arguments, we have found a magnetic
stress tensor in magnetized matter that, although manifestly
asymmetric, has the correct vacuum limit and appears to be
consistent with all conceptual tests that we have applied. For the
special case of a magnetized fluid, with magnetization parallel to
the macroscopic field, we show that the longitudinal pressure
remains negative and, unless the magnetization is extremely
strong, the perpendicular pressure remains positive. Thus, a
neutron star with an extremely strong magnetic field will
definitely {\it not} collapse to a {\it prolate} structure, as has
been proposed\cite{Herman+Hugo+Aurora+otro,Herman+Hugo+Aurora},
but, if anything, to an {\it oblate} structure, assuming that the
star's fluid pressure does not impede it.

\section*{Acknowledgments}
The authors thank the organizers of the {\it International
Workshop on Strong Magnetic Fields} for providing the stage for
several interesting discussions, and our fellow participants for
motivating us to do the present calculation and refine the
arguments presented here. The main calculation was carried out
over an excellent crab dinner at the {\it ``paladar''} Aries. (The
astronomical overtones of both the menu and the restaurant's name
are pure chance!) This work and our participation in the workshop
were supported by the workshop organizers, and by FONDECYT-Chile
through the regular research grant 1020840 and the L.C. grant
8000017.


\begin{thebibliography}{99}

\bibitem{Chaichian} M.~Chaichian~{\it et al.}, Phys.~Rev.~Lett.
84, 5261 (2000).

\bibitem{Herman+Hugo+Aurora+otro} R.~Gonz\'alez~Felipe~{\it et al.},
\emph{Quantum instability of magnetized stellar object},
astro-ph/0207150 (2002).

\bibitem{Herman+Hugo+Aurora} A. P\'erez Mart\'{\i}nez, H. P\'erez Rojas
and H.~J.~Mosquera Cuesta, Eur.~Phys.~J. {\bf C} DOI:
10.1140/epjc/s2003-01192-6 (2003), and these Proceedings.

\bibitem{Canuto}
V. Canuto, Phys.~Rev.~{\bf A}3, 648 (1971).

\bibitem{Easson}
I.~Easson and C.~J~.Pethick, Phys.~Rev.~{\bf D}16, 275 (1977).

\bibitem{Blandford}
R.D.\ Blandford and L.\ Hernquist, J.\ Phys.\ {\bf C}15, 6233 (1982).

\bibitem{Pethick}This is in agreement with C. J. Pethick, in
Structure and Evolution of Neutron Stars, eds. D. Pines, R.
Tamagaki, \& S. Tsuruta (Redwood City: Addison Wesley), p. 115
(1992).

\bibitem{Jackson}J. D. Jackson, Classical
Electrodynamics, 2nd. Edition, John Wiley \& Sons (1975).

\bibitem{Schutz}B. F. Schutz, A first course in general
relativity, Cambridge University Press (1985).

\end{thebibliography}
\end{document}